\documentclass[12pt,preprint,aps,pra]{revtex4}
\usepackage{graphicx}

\begin{document}

\title{Half-metallicity and giant magneto-optical Kerr effect in N-doped NaTaO$_3$}

\author{Y. Saeed$^{1}$, N. Singh$^{1, 2}$, and U. Schwingenschlgl$^{1}$}

\affiliation{$^{1}$Physical Science \& Engineering division, KAUST, Thuwal 23955-6900, Kingdom of Saudi Arabia}

\affiliation{$^{2}$Solar and Photovoltaic Energy Research Center, KAUST, Thuwal 23955-6900, Kingdom of Saudi Arabia}

\begin{abstract}
We employ density functional theory using the modified Becke-Johnson (mBJ) approach to investigate the electronic and magneto-optical properties
of N-doped NaTaO$_3$. The mBJ results reveal a half metallic nature of NaTaO$_2$N, in contrast to results obtained by the generalized gradient approximation. 
We find a giant polar Kerr rotation of 2.16$^{\circ}$ at 725 nm wave length (visible region), which is high as compared to other half metallic perovskites as well as to the prototypical half metal PtMnSb. 
\end{abstract}

\keywords{Density Functional Theory, MBJ, Ferromagnetic Half Metal, Magneto-optical properties}

\maketitle

\section{Introduction}

Room temperature ferromagnetism has been reported for different doped oxides such as C/N-doped ZnO \cite{Pan,Shen,yang}, TiO$_2$ \cite{YangDai-apl,YangDai-cpl,Tao}, SnO$_2$ \cite{Rahman,Xiao} and confirmed recently \cite{Nagare-zno,Bao-tio2,Hong-sio2}. Room temperature ferromagnetism with half metallicity is reported for N-doped SrTiO$_3$ and BaTiO$_3$ \cite{Liu-sto-n,Tan-bto-n}, in which the magnetic interactions between the nearest and next-nearest N dopants result in a strong ferromagnetic coupling \cite{Yang-2011-sto-n-ferro}.
Ferromagnetic half-metals have potential applications in spintronics devices \cite{Engen,Groot} and also show unusual magneto-optical effects due to a metallic state for one spin channel and an insulating state for the other. Yang \emph{et al}. have reported that a high concentration of N is required for achieving a magnetic long rang order in perovskite oxide \cite{Yang-2011-sto-n-ferro}.

The perovskite oxide NaTaO$_3$ (NTO) is a ferroelectric material with high permittivity and low dielectric loss, which suggests usage in microwave devices
\cite{Rabe,Geyer,Axelsson}. Several ab-initio calculations have been performed to described the electronic properties of bulk NTO \cite{Choi-2011} but a detailed study of the electronic structure and magneto-optical properties of N-doped NTO is missing in the literature. The magneto-optical Kerr effect of doped NTO
is interesting  for magneto-optical reading and recording devices \cite{Fiebig}. The N-doped perovskite oxide NTO is a 5$\emph{d}$ system. Therefore, electron correlation effects are expected to be small as compared to 3$\emph{d}$ systems such as SrTiO$_3$ and BaTiO$_3$. In the following we establish a half metallic nature for NaTaO$_{1-x}$N$_x$ ($x=0.04-0.33$) and discuss the electronic structure in comparison to the strongly correlated
perovskites SrTiO$_3$ and BaTiO$_3$. We also address the magneto-optical Kerr effect in N-doped NTO. 

\section{computational method}

Our calculations are based on density functional theory, using the full-potential linearized augmented plane wave approach as implemented in the WIEN2k code \cite{wien2k}. We use the modified Becke-Johnson (mBJ) exchange correlation potential \cite{mBJ}. The popular generalized gradient approximation (GGA) \cite{GGA-PBE} is employed to optimize the volume and the internal atomic coordinates. In general, the unit cell is divided into non-overlapping atomic spheres centered at the atomic sites and an interstitial region. The convergence parameter \textit{R}$_{mt}$\textit{K}$_{max}$, where \textit{K}$_{max}$ is the plane-wave cut-off and \textit{R}$_{mt}$ is the smallest muffin-tin radius, controls the size of the basis set. This convergence parameter is set to 7 together with \textit{G}$_{max}=24$. We use 66 \textbf{k}-points in the irreducible wedge of the Brillouin zone for calculating the electronic structure and a dense mesh of 480 \textbf{k}-points in the magneto-optical calculations. The cubic phase \emph{Pm$\overline{3}$m} ($a=b=c=3.93$ \AA\ and $\alpha=\beta=\gamma=90^\circ$) of NTO \cite{cubic-nto} is used in the present calculations for simplicity because the differences to the monoclinic phase $P2/m$ ($a=3.8995$ \AA, $b=3.8965$ \AA, and $c=3.8995$ \AA, $\alpha=\gamma=90^\circ$, and $\beta=90.15^\circ$) are subordinate \cite{monoclinc-nto}. As a consequence, the electronic band structures and density of states (DOS) are found to be very similar in both phases \cite{wang-JPCC2011,Lin-APL-phases}. 

\begin{figure}
\includegraphics[width=0.5\textwidth,clip]{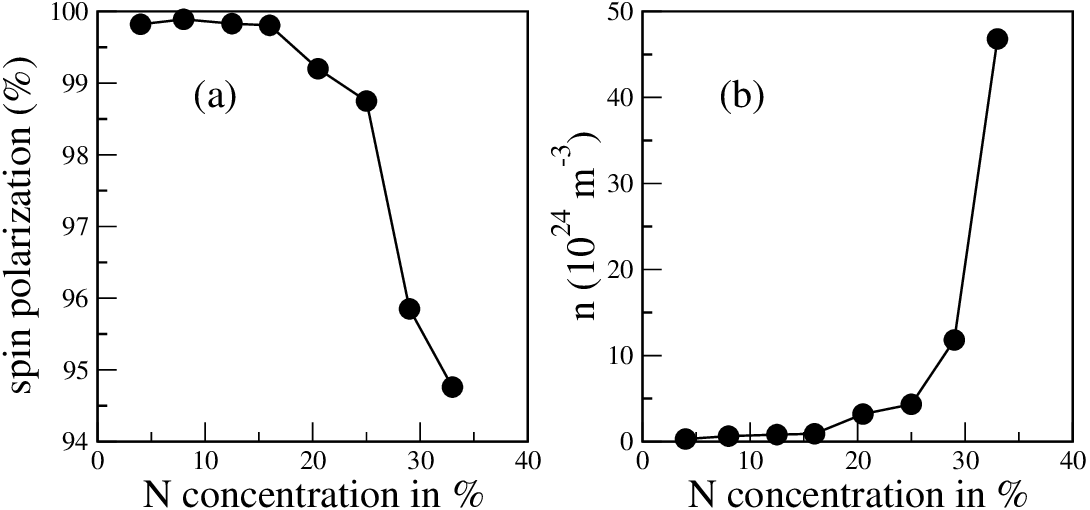}
\caption{Calculated spin polarization and majority spin hole density \emph{n} as a function of N-concentration, as obtained by the GGA+SOC approximation.}
\end{figure}

\section{Results and Discussion}

Our optimized lattice parameter (using the GGA) of cubic NTO is 3.98 \AA, which is in good agreement with the experimental value of 3.93 \AA\ \cite{cubic-nto}. We replace one O with one N to form the oxynitrate (NaTaO$_{2}$N). The optimized lattice parameters of NaTaO$_{2}$N is slightly increased to 4.03 \AA. In order to find the magnetic ground state, we construct a $1\times1\times2$ supercell using the optimized structure and replace two O atoms with N. We compare the ground state energies of ferromagnetic (FM) and anti-ferromagnetic (AFM) configurations. The magnetic energy $E_{mag}=E_{FM}-E_{AFM}=-51.3$ meV, and the N-N distance is $\sim$4 \AA\  with a total magnetic moment of 2 $\mu_{B}$ per cell (or 1 $\mu_{B}$ per N atom) in FM case. The Curie temperature
$T_{C}$ is calculated using the mean-field Heisenberg model, i.e., $T_{C}=(2/3)E_{mag}/k_{B}$ \cite{Kudrnovsk,Maca}. The 
calculated $T_{C}$ for NaTaO$_{2}$N is 396 K, which is close to that of N-doped SrTiO$_3$ and BaTiO$_3$ at the same N-N distance \cite{Yang-2011-sto-n-ferro}.
In order to observe the long range FM order, we study a high N-doping of 33\% by replacing one O by one N in a  single unit cell. A magnetic moment is induced as the delocalized N \emph{p} states become polarized, where 0.15 $\mu_{B}$ come from the interstitial, 0.13 $\mu_{B}$ come from O and 0.61 $\mu_{B}$ come from N, summing upto 1 $\mu_{B}$ per N atom. 

To explain the induced spin-polarization in NaTaO$_{2}$N, we analyzes DOS and electronic band structure obtained by GGA approximation (not shown here). The DOS shows a half-metallic character with a metallic state for the minority spin and an insulating state for the majority spin. To confirm the half-metallicity, we include spin orbit coupling (SOC) along with GGA in the calculations, finding that NaTaO$_{2}$N becomes a metal since the majority spin states crosses the Fermi level. The spin polarization ($=\frac{N\uparrow-N\downarrow}{N\uparrow+N\downarrow}$, where \textit{N} is the number of states at the Fermi level) of NaTaO$_{2}$N is obtained $\sim$94\%. In order to find the exact N-concentration at which the character of the system changes from a half-metal to metal, we construct a $3\times3\times3$ supercell and vary the N-concentration from 4\% to 33\% (including SOC in the calculations). In Fig.\ 1(a), we plot the spin-polarization as a function of the N-concentration. Below 16\% N-doped, NTO shows a $\sim$99.8\% spin-polarization which decreases sharply to $\sim94\%$ at 33\% N-doping. In Fig.\ 1(b), we plot the hole density (holes per volume) for the majority spin channel. Similar to the spin-polarization, the hole density increases rapidly upto 46.8$\times10^{24}$ m$^{-3}$ as the N-concentration increases to 33\%, while the hole density is almost constant for low N-concentration.

\begin{figure}
\includegraphics[width=0.5\textwidth,clip]{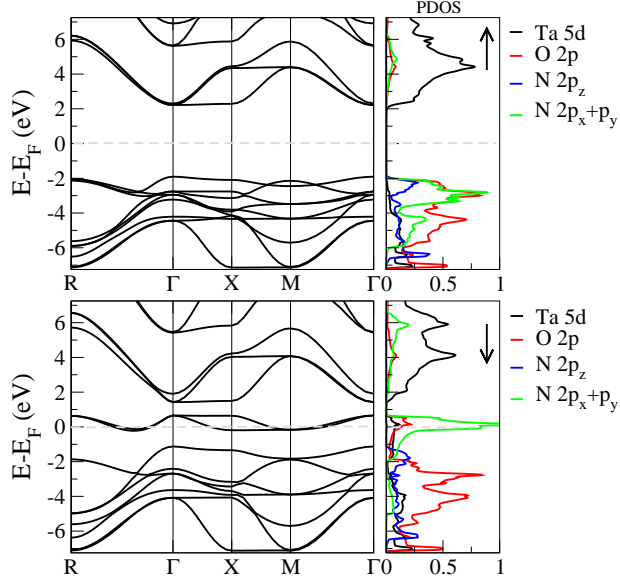}
\caption{Band structure and DOS of NaTaO$_{2}$N as obtained by the mBJ approximation.}
\end{figure}

Recently, Guo \emph{et al. }\cite{Guo-mBJLDA 2011} have applied the mBJ approach successfully to improve the half-metallic ferromagnetism in zincblende MnAs, which turns into a half-metal without affecting the \emph{d} \emph{t}$_{2g}$ bands. We apply the same method to NaTaO$_{2}$N. The calculated band structure and DOS in Fig.\ 2 show a truly half-metallic nature for NaTaO$_{2}$N. The majority spin bands are similar to pristine NTO with a gap of 3.96 eV, which is in excellent agreement with experiments \cite{Lin-APL-phases} and the previous GW calculations \cite{wang-JPCC2011}. The minority spin channel is metallic due to a non-zero DOS at the Fermi level. The band splitting at the Fermi level along R-$\Gamma$ and M-$\Gamma$ is very small, while along $\Gamma$-X-M, it is large. The calculated plasma frequency $\omega_{p}$ from the minority spin channel due to metallic nature, is 2.7 eV, which is smaller in the ferromagnetic half-metal PtMnSb ($\omega_{p}=4.5$ eV) \cite{Picozzi}, reflecting less dispersed bands. The calculated DOS shows that the valence bands (majority spin) are a combination of N 2\emph{p} and O 2\emph{p} states. The bottom of the conduction bands is composed of Ta 5\emph{d} states (see Fig.\ 2). For the minority spin channel, the N 2\emph{p} bands cross the Fermi level (with small O 2\emph{p} contributions). The N 2$p^{\uparrow\downarrow}$ states split into $(p_{x}+p_{y})^{\uparrow\downarrow}$ and $p_{z}^{\uparrow\downarrow}$ bands. There is no shifting of peak position with respect to energy is observed at the Fermi level in N 2$(p_{x}+p_{y})^{\uparrow\downarrow}$ and N 2$p_{z}^{\uparrow\downarrow}$ states from  the N-doped SrTiO$_3$ and BaTiO$_3$ \cite{Yang-2011-sto-n-ferro} where N 2$p_{y}+p_{z}$ and N 2$p_{x}$ have different peak position at the Fermi level. There is a strong hybridization between the N 2$p^{\uparrow\downarrow}$ and O 2$p^{\uparrow\downarrow}$ states for the minority spin channel. The Ta 5\emph{$d^{\uparrow\downarrow}$} bands do not change with the N-concentration.

\begin{figure}
\includegraphics[width=0.5\textwidth,clip]{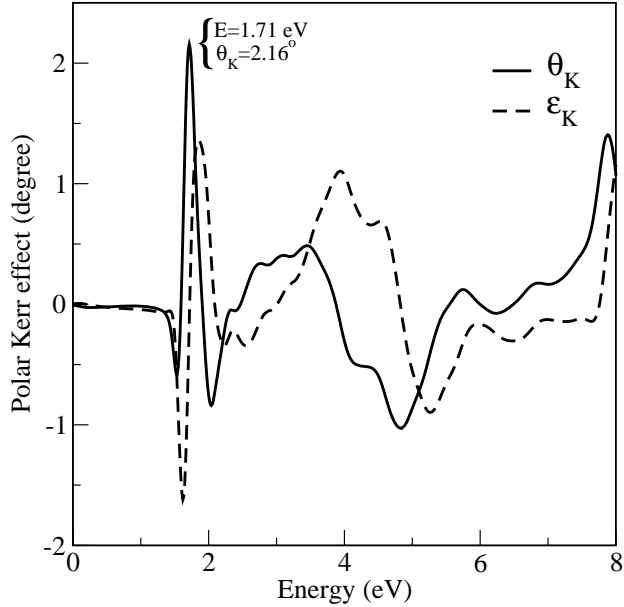}
\caption{Calculated polar Kerr angle $\theta_{K}$ and Kerr ellipticity $\varepsilon_{K}$ of NaTaO$_{2}$N.}
\end{figure}

Intense search aim at materials with large magneto-optical peaks in the low wave-length region to be used for high-density storage \cite{Grundy-high disk}. Both borates \cite{borates} and Zintl compounds \cite{Zintl,Zintl1}, can shows a remarkable Kerr signal in the low energy range. The Kerr rotation $\theta_{K}$ and Kerr ellipticity $\varepsilon_{K}$ of half metallic NaTaO$_{2}$N are shown in Fig.\ 3. We find a value of $\theta_{K}=2.16^{\circ}$ at 1.71 eV ($\sim$725 nm), which is higher than in BiNiO$_3$ ($\theta_{K}=1.28^{\circ}$) \cite{M. Q. Cai-BiNiO3} and the Heusler compound PtMnSb ($\theta_{K}=1.27^{\circ}$) \cite{van Engen-1983,Lobove-2012-ptmnsb}. 
The high Kerr angle is an intraband effect, and not due to the SOC (which creates an imbalance in the optical transitions in PtMnSb and NiMnSb \cite{van Ek-1997}, for example. 
For the minority spin channel, the band structure of NaTaO$_{2}$N shows a set of parallel bands across the Fermi level (R-$\Gamma$, $\Gamma$-X-M, and $\Gamma$-M) which consist of N 2{$(p_{x}+p_{y})^{\downarrow}$}
states. These parallel bands give rise to intraband transitions which contribute significantly to the Kerr spectrum in the low energy range. In NaTaO$_{2}$N, the separation between these bands is much smaller than in PtMnSb \cite{van Ek-1997}. This past explain the higher magneto-optical Kerr effect in NaTaO$_2$N. The calculated Kerr ellipticity $\varepsilon_{K}$ has a maximum of $\sim$1.7$^{\circ}$ at 1.6 eV.

\section{Conclusion}

In conclusion, we have presented first principles results of the band structure, DOS, and magneto-optical properties of N-doped NaTaO$_3$, as obtained from density functional theory. Our results for NaTaO$_{1-x}$N$_{x}$ ($x=0.04-0.33$) show that the GGA+SOC approach gives a 99\% spin-polarization at low N-concentrations upto 16\%. The mBJ+SOC approach results in a pure ferromagnetic half-metal in contrast to the GGA+SOC. We observe a giant magneto-optical Kerr signal of {$\theta_{K}$=2.16$^{\circ}$ at $\sim$725 nm in NaTaO$_2$N, which is the highest Kerr angle among the ferromagnetic half-metals in UV-visible region. The origin of the high Kerr angle is attributed to intraband transitions involving the N 2\textbf{$(p_{x}+p_{y})^{\downarrow}$} orbital due to parallel bands around the Fermi level. The large Kerr rotation in NaTaO$_2$N in the visible region may find applications in red/infrared laser magneto-optical devices and the half metallic nature of NaTaO$_2$N is interesting for spintronics devices.


\begin{thebibliography}{10}
\bibitem{Pan}H. Pan, J. B. Yi, L. Shen, R. Q. Wu, J. H. Yang, J. Lin, Y. P. Feng, J. Ding, L.H. Van, and J. H. Yin, Phys. Rev. Lett. 99, 127201 (2007).

\bibitem{Shen}L. Shen, R. Q. Wu, H. Pan, G. W. Peng, M. Yang, Z. D. Sha, and Y. P. Feng, Phys. Rev. B 78, 073306 (2008).

\bibitem{yang}K. Yang, R. Wu, L. Shen, Y. P. Feng, Y. Dai, and B. Huang, Phys. Rev. B 81, 125211 (2010).

\bibitem{YangDai-apl}K. Yang, Y. Dai, B. Huang, and M.-H. Whangbo, Appl. Phys. Lett. 93, 132507 (2008). 

\bibitem{YangDai-cpl}K. Yang, Y. Dai, B. Huang, and M.-H. Whangbo, Chem. Phys. Lett 481, 99 (2009). 

\bibitem{Tao}J. G. Tao, L. X. Guan, J. S. Pan, C. H. A. Huan, L.Wang, J. L. Kuo, Z. Zhang, J. W. Chai, and S. J. Wang, Appl. Phys. Lett. 95, 062505 (2009).

\bibitem{Rahman}G. Rahman and V. M. Garc\'{i}a-Su\'{a}rez, Appl. Phys. Lett. 96, 052508 (2010).

\bibitem{Xiao}W.-Z. Xiao, L.-L. Wang, L. Xua, Q. Wan, and B. S. Zou, Solid State Commun. 149, 1304 (2009).

\bibitem{Nagare-zno} B. J. Nagare, S. Chack, and D. G. Kanhere, J. Phys. Chem. A 114, 2689 (2010).

\bibitem{Bao-tio2}N. N. Bao, H. M. Fan, J. Ding, and J. B. Yi, J. Appl. Phys. 109, 07C302 (2011). 

\bibitem{Hong-sio2}N. H. Hong, J.-H. Song, A. T. Raghavender, T. Asaeda, and M. Kurisu, Appl. Phys. Lett. 99, 052505 (2011).

\bibitem{Liu-sto-n}C. M. Liu, X. Xiang, and X. T. Zu, Chin. J. Phys. 47, 893 (2009).

\bibitem{Tan-bto-n}X. Tan, C. Chen, K. Jin, and B. Luo, J. Alloy. Compd. 509, L311 (2011).

\bibitem{Yang-2011-sto-n-ferro}K. Yang, Y. Dai, and B. Huang, Appl. Phys. Lett. 100, 062409 (2012).

\bibitem{Engen}P. G. van Engen, K. H. J. Buschow, and R. Jongebreur, Appl. Phys. Lett. 42, 202 (1982). 

\bibitem{Groot}R. A. de Groot, F. M. Mueller, P. G. van Engen, and K. H. J. Buschow, Phys. Rev. Lett. 50, 2024 (1983).

\bibitem{Rabe}K. Rabe, C. H. Ahn, and J.-M. Triscone, Physics of Ferroelectrics: A Modern Perspective, Topics in Applied Physics (Springer, Berlin, 2007), Vol. 105. 

\bibitem{Geyer}R. G. Geyer, B. Riddle, J. Krupka, and L. A. Boatner, J. Appl. Phys. 97, 104111 (2005). 

\bibitem{Axelsson}A.-K. Axelsson, Y. Pan, M. Valant, and N. Alford, J. Am. Ceram. Soc.\textbf{ }92, 1773 (2009).

\bibitem{Choi-2011}M. Choi, F. Oba, and I. Tanaka, Phys. Rev. B 83, 214107 (2011). 

\bibitem{Fiebig} M. Fiebig, J. Phys. D: Appl. Phys. 38, R123 (2005).

\bibitem{wien2k}P. Blaha, K. Schwarz, G. Madsen, D. Kvasicka, and J. Luitz, WIEN2k, An Augmented Plane Wave + Local Orbitals Program for Calculating Crystal Properties (TU Vienna, Vienna, 2001).

\bibitem{mBJ}F. Tran and P. Blaha, Phys. Rev. Lett. 102, 226401 (2009).

\bibitem{GGA-PBE}J. P. Perdew, K. Burke, and M. Ernzerhof, Phys. Rev. Lett. 77, 3865 (1996). 

\bibitem{cubic-nto}International Center for Diffraction Data, JCPDS Card No. 742488 (2001).

\bibitem{monoclinc-nto}International Center for Diffraction Data, JCPDS Card No. 742479 (2001).

\bibitem{wang-JPCC2011}H. Wang, F. Wu, and H. Jiang, J. Phys. Chem. C 115, 16180 (2011).

\bibitem{Lin-APL-phases}W. H. Lin, C. Cheng, C. C. Hu, and H. Teng, Appl. Phys. Lett. 89, 211904 (2006).

\bibitem{Kudrnovsk}J. Kudrnovsk\'{y}, I. Turek, V. Drchal, F. M\'{a}ca, P. Weinberger, and P. Bruno, Phys. Rev. B 69, 115208 (2004).

\bibitem{Maca}F. M\'{a}ca, J. Kudrnovs\'{k}, V. Drchal, and G. Bouzerar, Appl. Phys. Lett. 92, 212503 (2008).

\bibitem{Guo-mBJLDA 2011}S.-D. Guo and B.-G. Liu, Euro. Phys. Lett. 93, 47006 (2011). 

\bibitem{Picozzi}S. Picozzi, A. Continenza, and A. J. Freeman, J. Phys. D: Appl. Phys. 39, 851 (2006).

\bibitem{Grundy-high disk}P. J. Grundy, in Electronic and Magnetic Properties of Metals and Ceramics, Materials Science and Technology, edited by K. H. J. Buschow (VCH, 1994), Vol. 3B, p. 575.

\bibitem{borates}Y. Saeed, N. Singh, and U. Schwingenschl\"ogl, J. Appl. Phys. 110, 103512 (2011).

\bibitem{Zintl}N. Singh and U. Schwingenschl\"ogl, Chem. Phys. Lett. 508, 29 (2011).

\bibitem{Zintl1}N. Singh and U. Schwingenschl\"ogl, Appl. Phys. Lett. 100, 151906 (2012).

\bibitem{M. Q. Cai-BiNiO3} M. Q. Cai, X. Tan, G. W. Yang, L. Q. Wen, L. L. Wang, W. Y. Hu, and Y. G. Wang, J. Phys. Chem. C 112, 16638 (2008).

\bibitem{van Engen-1983}P. G. van Engen, K. H. J. Buschow, R. Jongebreur, and M. Erman, Appl. Phys. Lett. 42, 202 (1983).

\bibitem{Lobove-2012-ptmnsb}I. D. Lobov, A. A. Makhnev, and M. M. Kirillova, Phys. Met. Metallogr. 113, 135 (2012), and references therein.

\bibitem{van Ek-1997}J. van Ek and J. M. Maclaren, Phys. Rev. B 56, R2924 (1997).

\end{thebibliography}
\end{document}